\documentclass[epj,nopacs]{svjour}
\usepackage{graphicx}
\usepackage[english]{babel}
\usepackage{amsmath}

\begin{document}

\title{Constraining Spin-One Color-Octet Resonances Using CDF and ATLAS Data}

\author{Alfonso R. Zerwekh\thanks{\email{alfonsozerwekh@uach.cl}}}

\institute{Instituto de F\'isica, Facultad de Ciencias,
 Universidad Austral de Chile
and Centro Cient\'ifico-Tecnol\'ogico de Valpara\'iso 
 Casilla 567, Valdivia, Chile}

\date{}

\abstract{
In this paper, we study the production of spin-one color-octet
resonances (colorons) at hadron colliders in a model independent
way. We use dijets data measured by CDF (at $\sqrt{  s}=1.96$ TeV and
${\cal L}=1.13$ pb$^{-1}$) and ATLAS (at $\sqrt{  s}=7$ TeV and
${\cal L}=315$ nb$^{-1}$) collaborations at the 
Tevatron and the LHC respetively to impose limits on the coupling of
colorons to fermions. We show that CDF data still produce the more
stringent limits on the coloron coupling constant.  
}

\maketitle 
\section{Introduction}

Hadron colliders are the right tools for exploring new energy
scales searching for new Physics. During the last decades this role
has been played by the Tevatron. The discovery of the top
quark \cite{Abe:1995hr,Abachi:1994td}, with
its originally unexpected huge mass is only one example of its
significant contribution to our understanding of Particle Physics. Now
the LHC is running and we hope that it will uncover phenomena beyond
the Standard Model. The observation of New Physics is not, however,
the only contribution of a collider. While new particles escape
detection, the experimental data can be used to constrain
models. Theorists have been extremely creative proposing extensions
of the Standard Model and only direct or indirect experimental
information can guide us to the path chosen by Nature.

Many proposed models predicts the existence of spin-one color-octet
particles. They have been studied, for example, in non-minimal Technicolor
(color-octet technirho,
$\rho_8$)
\cite{Eichten:1984eu,Lane:1991qh,Zerwekh:2001uq,Zerwekh:2004vy,Zerwekh:2006te},
Topcolor (coloron) \cite{Chivukula:1996yr,Simmons:1996fz}, models with
extra dimensions (Kaluza-Klein excitation of the
gluon) \cite{Lillie:2007yh} and chiral 
color models (axigluon)
\cite{Frampton:1987ut,Frampton:1987dn,Ferrario:2009bz,Zerwekh:2009vi}  
 but also in a model independent way
 \cite{Dobrescu:2007yp,Kilic:2008pm,Kilic:2009mi}. It is 
generally expected that this kind of particles, which we will call
generically ``colorons'', if they
exists, must be copiously produced at hadron colliders due to their
color charge and they must appear as resonances in the dijet
spectrum. Now, we dispose of the first data of the ATLAS
search of resonances that can decay into dijets \cite{Collaboration:2010bc}
 and we felt it is
pertinent to upgrade our studies on spin-one color-octet resonaces and
compare the limits obtained from ATLAS data with the ones obtained
from CDF \cite{Aaltonen:2008dn} measurements. 

In section \ref{theory} we recall a general description of colorons
based exclusively on gauge symmetry and we show that a spin-one
color-octet resonance cannot couple to two gluons by a dimension four
operator and that the coupling constant of the coloron to quarks is
not completely fixed by gauge symmetry.

In section \ref{results} we show the limits imposed by experimental
data on the coloron coupling constant and we conclude that the more stringent
limits are still obtained from CDF measurements. 

Finally, in section \ref{conclusions}, we discuss our results.

\section{Theoretical Description}\label{theory}
We start considering two vector fields($A_{1\mu}^a$ and $A_{1\mu}^a$)
in the adjoint representation of the color group $SU(3)_c$ and we work
in a basis where both fields transform like gauge fields. With these
ingredients we can write down the following gauge invariant
Lagrangian:

\begin{equation}
\mathcal{L}=-\frac{1}{4} {F}_{1\mu\nu}^{a} {F}_{1}^{a\mu\nu}-\frac{1}{4} {F}_{2\mu\nu}^{a} {F}_{2}^{a\mu\nu}+\frac{M^{2}}{2 {g}_{2}^2}\left( {g}_{1}A_{1\mu}^{a}- {g}_{2}A_{2\mu}^{a}\right)^{2}
\label{eq:N2} 
\end{equation} 
where $M$ is a new mass scale present in the model. 
At this point it is necessary to remark that the mass term introduced
in our Lagrangian is completely consistent with gauge invariance
since, as it is straightforward to prove, the linear combination $(
{g}_{1}A_{1\mu}^{a}- {g}_{2}A_{2\mu}^{a})$  transforms homogeneously
under local gauge
transformations\cite{Zerwekh:2001uq,Zerwekh:2006te,Zerwekh:2003zz} . 

The mass matrix originated by this Lagrangian (which is in this case
the most general mass matrix for the gauge sector compatible with
gauge symmetry) is exactly diagonalizable. In fact, the mass
eigenstates fields can be written as

\begin{eqnarray}
G & = & A_{1}\cos(\alpha)+A_{2}\sin(\alpha)\label{eq:EigenVector1}\\
V & = & -A_{1}\sin(\alpha)+A_{2}\cos(\alpha)\label{eq:EigenVector2}
\end{eqnarray}
Where, for simplicity, we have dropped the Lorentz and color indexes
and we have defined $\sin(\alpha)=g/ {g}_{2}$ and $g$ is: 

\[
g=\frac{ {g}_{1} {g_{2}}}{\sqrt{ {g}_{1}^{2}+ {g}_{2}^{2}}}.
\]

The masses of the physical states are:

\begin{eqnarray*}
m_{G} & = & 0\\
m_{V} & = & \frac{M}{cos(\alpha)}
\end{eqnarray*}
\noindent
so, we see that the Lagrangian describes, in fact, a massless gauge
boson and a massive spin-one resonance in the adjoint
representation. Clearly, we identify $G$ with the physical gluon
and the massive state $V$ is the coloron

Let us study, for a moment, the decay process $V\rightarrow G G$.
The relevant part of the Lagrangian, written in terms of the physical
fields, is:

\begin{eqnarray}
\mathcal{L}_{VGG} & = &
\left(-g_{1}\cos^{2}(\alpha)\sin(\alpha)+g_{2}\cos(\alpha)\sin^{2}(\alpha)\right)
f^{abc}\cdot \notag\\ 
& \cdot & \{\partial_{\mu}G_{\nu}^{a}G^{b\mu}V^{c\nu}+
+\partial_{\mu}G_{\nu}^{a}V^{b\mu}G^{c\nu}+\partial_{\mu}V_{\nu}^{a}G^{b\mu}G^{c\nu}\}
\notag \\
 & &
\label{eq:L211} 
\end{eqnarray}

It is clear that, due to the definition of $\alpha$,
$ {g}_{1}\cos(\alpha)= {g}_{2}\sin(\alpha)$. 
Using this identity, we can see that the coupling constant of the
$V G G$ interactions vanishes exactly
\cite{Simmons:1996fz,Zerwekh:2001uq,Kilic:2008pm}. This result is an 
important consequence of the gauge symmetry of the model. It is
necessary to say, however, that this coupling can be restored by
dimension 6 operators \cite{Kilic:2008pm,Chivukula:2001gv}. Nevertheless we will
not consider here this case.   

Let us turn our attention to the coupling with quarks. Because we have
now two fields that transform as gauge fields ($A_{1\mu}^a$ and
$A_{2\mu}^a$) any combination of the form:

\begin{equation}\label{DirectCoupling}
 {g}_{1}(1-k)A_{1\mu}^a+ {g}_{2}kA_{2\mu}^a
\end{equation}
\noindent
where $k$ is
an  arbitrary parameter, can be used to construct a covariant
derivative \cite{Zerwekh:2006te,Zerwekh:2003zz}.
 Notice that the resulting generalized covariant derivative gives
 origin to  a direct coupling, between the field $A_{2\mu}$ and
the fermions, parameterized by the new constant $k$. In other words, now
$A_{2\mu}$ couples to fermions not only through its mixing with
$A_{1\mu}$ (parameterized by the angle $\alpha$) but also through the
new term introduced into the covariant derivative. Previously, we have 
used this kind of direct coupling in order to study the phenomenology
of the color-octet technirho \cite{Zerwekh:2006te}, it has been also 
useful in allowing us to propose a mechanism for reducing axigluon
couplings \cite{Zerwekh:2009vi} and recently this kind of terms played
a crucial role in the construction of a model with two composite Higgs
bosons \cite{Zerwekh:2009yu}. Interestingly, this theoretical
construction, \textit{i.e.} the introduction of a linear combination
of vector fields like (\ref{DirectCoupling}) which depends on the
arbitrary parameter $k$, 
naturally arises in Deconstruction theories with ``delocalized''
fermions \cite{Chivukula:2005bn,Chivukula:2005xm}.

As a consequence of the generalized covariant derivative, the coupling of
the physical (mass eigenstates) vector bosons to quarks can be written as:
\begin{eqnarray}
\mathcal{L}_{V\bar{q}q}&=&g_{QCD}\frac{\lambda^a}{2}\bar{\psi}G_{\mu}^a\gamma^{\mu}\psi\notag \\
&+&g_{QCD}\tan\alpha\left(\frac{k}{\sin^2\alpha}-1\right)
\frac{\lambda^a}{2}\bar{\psi}V_{\mu}^a\gamma^{\mu}\psi
\end{eqnarray}    
where $g_{QCD}=g/\sqrt{2}$. Notice that the gluon couples to quarks in
the usual way while the coloron coupling constant
($g_V=g_{QCD}\tan\alpha(k/\sin^2\alpha -1)$)
depends not only on the mixing angle but also on the arbitrary
parameter $k$ and hence its value is not fixed by the gauge
principle. This fact allows us to try to use experimental data in
order to constrain the value of $g_V$ and make the coloron invisible
to present searching efforts.

\section{Results}\label{results}

We implemented the model described in the previous section into
CalcHEP \cite{Pukhov:1999gg,Pukhov:2004ca}
 and use it to generate events and compute cross sections for
the coloron production and decay into two jets which
could be compared to experimental data. Notice that, since the
coloron does not couple to two gluons, as was shown above, we only have
quarks in the initial and final states. Two set of events were
generated: the first one using $\sqrt{  s}=7.0$ TeV and the
following set of cuts for comparing with
ATLAS data:

\begin{eqnarray*}
  p_T^{j_1}&>&80.0 \mathrm{ GeV}\\
p_T^{j_2}&>&30.0 \mathrm{ GeV}\\
|\eta_i|<1.3 & \mathrm{ or } & 1.8<|\eta_i|<2.5
\end{eqnarray*}
where $\eta_i$ is the pseudo-rapidity of jet $i$ and $j_1$ represents
the jet with higher transversal momentum. For the second one, we use $\sqrt{  s}=1.96$ TeV and
$|\eta_i|<1.0$ in order to be compared with CDF data. For all the computations
we use the CTEQ6l parton distribution function \cite{Pumplin:2002vw}
. To obtain our limits we
vary both the coloron mass ($M_V$) and the coloron coupling constant
($g_V$) in the intervals $M_V \in [400,1500]$ GeV and $g_V \in
[0.05,2.50]$. 

\begin{figure}
\begin{centering}
\includegraphics[scale=0.20]{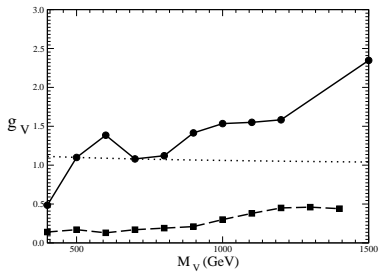}
\par\end{centering}

\caption{Upper limits on  the coloron coupling constant as a function
  of coloron Mass obtained by demanding escape detection by ATLAS 
  (dots with continuous line) and CDF (squares and dashed line). As a
reference, we include in the figure the value of the (running) QCD
coupling constant (dotted line).} 
\label{Flo:Limites}
\end{figure}

Our results are shown in figure \ref{Flo:Limites}. The dots with
continuous line represent the maximum value of $g_V$ compatible with
the $95 \%$ C.L. upper limits obtained by ATLAS at $\sqrt{  s}=7.0$ TeV
and ${\cal L}=315$ nb$^{-1}$. On the other hand, the squares with
dashed line is similar but was obtained by comparing our predicted
cross sections with the $95 \%$ C.L. upper limits reported by CDF
for $\sqrt{  s}=1.96$ TeV and ${\cal L}=1.13$ pb$^{-1}$. As a
reference, we include in the figure the value of the (running) QCD
coupling constant (dotted line).

As we can see, a spin-one color-octet resonance with QCD coupling
(that is $g_V=g_{QCD}$) is excluded by CDF but not by present ATLAS
data (except by ATLAS in a small region near $M_V=400$ GeV). In brief,
our results show that CDF data still produce the more stringent
limits on $g_V$.

It is worth to be remarked that despite the fact that our theoretical
description, which justify to set limits on $g_V$ since its value is
not fixed, was done for a color-octet field that has a vector
coupling to quarks, a similar construction can be done in the case of
chiral color models \cite{Zerwekh:2009vi}, and  in principle our
limits are also applicable to the axigluon.

\section{Conclusions}\label{conclusions}

In this work we have studied the production of color-octet spin-one
resonances at the Tevatron and the LHC and its decay in dijets. We
have compared our predicted cross section for several values of $g_V$
and $M_V$ with available upper limits from CDF and ATLAS
measurements. In this form, we were able to set up upper limits on the
value of $g_V$ in order to make those resonances invisible for the
above cited experiments. We find that the most stringent limits still
come from CDF data. Two reasons seems to determine this
result. First, as we have shown, this kind of resonance is only
produced by quark-anti-quark initial states. Being the Tevatron a $p
\bar{p}$ collider the production of a coloron (or even an axigluon)
would be favored in this collider. In other words, the Tevatron would
be a more propitious environment for searching color-octet spin-one
resonances. On the other, the ATLAS data used still reflect a low level
of luminosity (${\cal L}=315$ nb$^{-1}$). We hope that in the near
future, when data with higher integrated luminosity become available,
those limits can be improved. 

Finally, we wish to emphasize that, as was pointed out
elsewhere \cite{Zerwekh:2006te,Dobrescu:2007yp}, the LHC offers the
unique opportunity of studying the 
production of a pair of colorons. This process is theoretically
cleaner (almost model independent) and phenomenologically interesting.          

\section*{Acknowledgment}
 This work is partially supported by Fondecyt grant 1070880 and by the
 Conicyt grant ``Southern Theoretical Physics Laboratory'' ACT-91. TGD


\begin{thebibliography}{99}
\bibitem{Abe:1995hr}
  F.~Abe {\it et al.}  [CDF Collaboration],
  ``Observation of top quark production in $\bar{p}p$ collisions,''
  Phys.\ Rev.\ Lett.\  {\bf 74} (1995) 2626
  [arXiv:hep-ex/9503002].

\bibitem{Abachi:1994td}
  S.~Abachi {\it et al.}  [D0 Collaboration],
  ``Search for high mass top quark production in $p\bar{p}$ collisions at
  $\sqrt{s} = 1.8$ TeV,''
  Phys.\ Rev.\ Lett.\  {\bf 74} (1995) 2422
  [arXiv:hep-ex/9411001].

\bibitem{Eichten:1984eu}
  E.~Eichten, I.~Hinchliffe, K.~D.~Lane and C.~Quigg,
  ``Super Collider Physics,''
  Rev.\ Mod.\ Phys.\  {\bf 56}, 579 (1984)
  [Addendum-ibid.\  {\bf 58}, 1065 (1986)].

\bibitem{Lane:1991qh}
  K.~D.~Lane and M.~V.~Ramana,
  ``Walking technicolor signatures at hadron colliders,''
  Phys.\ Rev.\  D {\bf 44}, 2678 (1991).

\bibitem{Zerwekh:2001uq}
  A.~R.~Zerwekh and R.~Rosenfeld,
  ``Gauge invariance, color-octet vector resonances and double technieta
  production at the Tevatron,''
  Phys.\ Lett.\  B {\bf 503} (2001) 325
  [arXiv:hep-ph/0103159].

\bibitem{Zerwekh:2004vy}
  A.~R.~Zerwekh,
  ``Dijets Production In Nonminimal Technicolor And Limits On The Color-Octet
  Technirho Mass,''
  Int.\ J.\ Mod.\ Phys.\  A {\bf 19} (2004) 4387.

\bibitem{Zerwekh:2006te}
  A.~R.~Zerwekh,
  ``Consequences of partial vector meson dominance for the phenomenology of
  colored technihadrons,''
  Eur.\ Phys.\ J.\  C {\bf 49} (2007) 1077
  [arXiv:hep-ph/0603094].

\bibitem{Chivukula:1996yr}
  R.~S.~Chivukula, A.~G.~Cohen and E.~H.~Simmons,
  ``New strong interactions at the Tevatron?,''
  Phys.\ Lett.\  B {\bf 380}, 92 (1996)
  [arXiv:hep-ph/9603311].

\bibitem{Simmons:1996fz}
  E.~H.~Simmons,
  ``Coloron phenomenology,''
  Phys.\ Rev.\  D {\bf 55}, 1678 (1997)
  [arXiv:hep-ph/9608269].


\bibitem{Lillie:2007yh}
  B.~Lillie, L.~Randall and L.~T.~Wang,
  ``The Bulk RS KK-gluon at the LHC,''
  JHEP {\bf 0709} (2007) 074
  [arXiv:hep-ph/0701166].

\bibitem{Frampton:1987ut}
  P.~H.~Frampton and S.~L.~Glashow,
  ``Unifiable Chiral Color With Natural Gim Mechanism,''
  Phys.\ Rev.\ Lett.\  {\bf 58}, 2168 (1987).

\bibitem{Frampton:1987dn}
  P.~H.~Frampton and S.~L.~Glashow,
  ``Chiral Color: An Alternative to the Standard Model,''
  Phys.\ Lett.\  B {\bf 190}, 157 (1987).

\bibitem{Ferrario:2009bz}
  P.~Ferrario and G.~Rodrigo,
  Phys.\ Rev.\  D {\bf 80}, 051701 (2009)
  [arXiv:0906.5541 [hep-ph]].

\bibitem{Zerwekh:2009vi}
  A.~R.~Zerwekh,
  ``Axigluon Couplings in the Presence of Extra Color-Octet Spin-One Fields,''
  Eur.\ Phys.\ J.\  C {\bf 65} (2010) 543
  [arXiv:0908.3116 [hep-ph]].

\bibitem{Dobrescu:2007yp}
  B.~A.~Dobrescu, K.~Kong and R.~Mahbubani,
  ``Massive color-octet bosons and pairs of resonances at hadron colliders,''
  Phys.\ Lett.\  B {\bf 670} (2008) 119
  [arXiv:0709.2378 [hep-ph]].

\bibitem{Kilic:2008pm}
  C.~Kilic, T.~Okui and R.~Sundrum,
  JHEP {\bf 0807} (2008) 038
  [arXiv:0802.2568 [hep-ph]].

\bibitem{Kilic:2009mi}
  C.~Kilic, T.~Okui and R.~Sundrum,
  JHEP {\bf 1002} (2010) 018
  [arXiv:0906.0577 [hep-ph]].


\bibitem{Collaboration:2010bc}
  T.~A.~Collaboration,
  ``Search for New Particles in Two-Jet Final States in 7 TeV Proton-Proton
  Collisions with the ATLAS Detector at the LHC,''
  arXiv:1008.2461 [hep-ex].

\bibitem{Aaltonen:2008dn}
  T.~Aaltonen {\it et al.}  [CDF Collaboration],
   ``Search for new particles decaying into dijets in proton-antiproton
  Phys.\ Rev.\  D {\bf 79}, 112002 (2009)
  [arXiv:0812.4036 [hep-ex]].


\bibitem{Chivukula:2001gv}
  R.~S.~Chivukula, A.~Grant and E.~H.~Simmons,
   ``Two gluon coupling and collider phenomenology of color octet technirho
  mesons,''
  Phys.\ Lett.\  B {\bf 521}, 239 (2001)
  [arXiv:hep-ph/0109029].

\bibitem{Zerwekh:2003zz}
  A.~R.~Zerwekh,
  ``Effective description of a gauge field and a tower of massive vector
  resonances,''
  arXiv:hep-ph/0307130.

\bibitem{Zerwekh:2009yu}
  A.~R.~Zerwekh,
  Mod.\ Phys.\ Lett.\  A {\bf 25} (2010) 423
  [arXiv:0907.4690 [hep-ph]].

\bibitem{Chivukula:2005bn}
  R.~S.~Chivukula, E.~H.~Simmons, H.~J.~He, M.~Kurachi and M.~Tanabashi,
  Phys.\ Rev.\  D {\bf 71} (2005) 115001
  [arXiv:hep-ph/0502162].


\bibitem{Chivukula:2005xm}
  R.~S.~Chivukula, E.~H.~Simmons, H.~J.~He, M.~Kurachi and M.~Tanabashi,
  Phys.\ Rev.\  D {\bf 72} (2005) 015008
  [arXiv:hep-ph/0504114].

\bibitem{Pukhov:1999gg}
  A.~Pukhov {\it et al.},
  ``CompHEP: A package for evaluation of Feynman diagrams and integration  over
  multi-particle phase space. User's manual for version 33,''
  arXiv:hep-ph/9908288.

\bibitem{Pukhov:2004ca}
  A.~Pukhov,
  ``Calchep 2.3: MSSM, structure functions, event generation, 1, and generation
  of matrix elements for other packages,''
  arXiv:hep-ph/0412191.

\bibitem{Pumplin:2002vw}
  J.~Pumplin, D.~R.~Stump, J.~Huston, H.~L.~Lai, P.~M.~Nadolsky and W.~K.~Tung,
  ``New generation of parton distributions with uncertainties from global QCD
  analysis,''
  JHEP {\bf 0207} (2002) 012
  [arXiv:hep-ph/0201195].


 \end{thebibliography}
\end{document}